\documentclass[11pt]{article}
\usepackage[]{acl}
\usepackage[utf8]{inputenc}
\usepackage[protrusion=true,expansion=false]{microtype}
\usepackage{url}
\usepackage{booktabs}
\usepackage{amsmath}
\usepackage{amssymb}
\usepackage{graphicx}
\usepackage{xcolor}
\usepackage{enumitem}
\usepackage{tabularx}
\usepackage{multirow}
\usepackage{tikz}
\usetikzlibrary{arrows.meta,positioning,fit,backgrounds,calc,shapes.geometric}
\usepackage{balance}\setlist{nosep,leftmargin=1.2em}

\definecolor{inputc}{RGB}{70,110,160}
\definecolor{orchc}{RGB}{200,90,60}
\definecolor{knowc}{RGB}{90,140,90}
\definecolor{humanc}{RGB}{150,100,160}
\definecolor{outc}{RGB}{180,150,60}
\definecolor{bggray}{RGB}{245,245,245}
\definecolor{cmark}{RGB}{40,130,60}
\definecolor{xmark}{RGB}{170,50,40}
\definecolor{pmark}{RGB}{190,140,30}

\newcommand{\yes}{\textcolor{cmark}{\textbf{$\checkmark$}}}
\newcommand{\no}{\textcolor{xmark}{\textbf{$\times$}}}
\newcommand{\parmark}{\textcolor{pmark}{\textbf{$\sim$}}}

\title{The Orchestration Gap: Why Process Automation\\Stalls in Operationally Complex Industries}

\author{
  \normalfont
  \textbf{Jiechao Gao}\textsuperscript{1} \quad
  \textbf{Yuandong Pan}\textsuperscript{1}\thanks{\ \,Corresponding author.} \quad
  \textbf{Yuangang Li}\textsuperscript{2} \\[2pt]
  \textbf{Jie Wang}\textsuperscript{1} \quad
  \textbf{Kincho Law}\textsuperscript{1} \quad
  \textbf{Michael Lepech}\textsuperscript{1} \\[5pt]
  \textsuperscript{1}Stanford University \qquad
  \textsuperscript{2}University of California, Irvine \\[3pt]
  \texttt{\{jiechao, ydpan, jiewang, law, mlepech\}@stanford.edu} \quad
  \texttt{yuanganl@uci.edu}
}
\begin{document}

\maketitle

\begin{abstract}
Agentic systems have advanced quickly on digitally native tasks, yet they have barely touched the industries where coordinated automation could matter most: logistics, healthcare operations, construction, and the many sectors whose work is spread across incompatible tools and many hands. We argue that the reason is a missing abstraction. The value in these settings does not come from a single capable model invocation; it comes from \emph{orchestration}, the runtime that coordinates multi-step workflows, enforces hard domain constraints, manages human approval, and bridges legacy systems. We develop this idea into a usable conceptual frame. We give an operational test for which workflows are orchestration-bound, a decomposition that separates how tangled a workflow is from how much of its effort is coordination and what that coordination is worth, and a feature-level account of why today's multi-agent frameworks leave a specific gap. We then advance our central claim: the right automation path is staged, and which architectural guarantee carries the most weight depends on a sector's dominant source of friction. Constraint enforcement is load-bearing under regulatory friction; explainability is load-bearing under liability friction. We close with the research program this view implies.
\end{abstract}

\section{Introduction}

Industrial AI has advanced fastest where the environment is already digital. Software engineering, financial trading, and customer support offer clean interfaces and structured data, and agentic systems have made real progress there. The industries where coordinated automation could deliver the most operational value look nothing like this. Their workflows are fragmented across many tools and stakeholders, their decisions lean on uncodified experience, and their digital infrastructure consists of disconnected legacy systems that were never designed to interoperate.

\begin{figure}[t]
\centering
\includegraphics[width=\columnwidth]{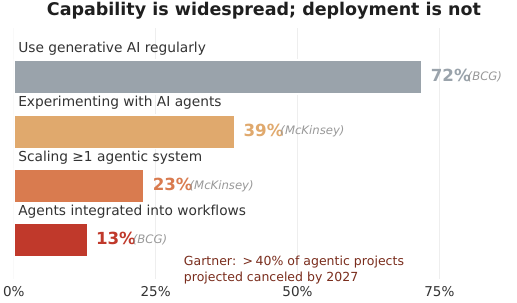}
\caption{The deployment gap that motivates this paper. Generative tools are in routine use and experimentation with agents is common, but the share of organizations that have moved agents into integrated, production workflows is small, and a large fraction of agentic projects is projected to be abandoned. Reported figures are from the cited industry surveys \citep{bcg2025aiatwork,mckinsey2025stateofai,gartner2025agentic}; the surveys differ in population and wording, so the bars are indicative of the gap rather than directly comparable points.}
\label{fig:gap}
\end{figure}

The symptom is visible in deployment statistics (Figure~\ref{fig:gap}). McKinsey's 2025 survey found 23\% of organizations scaling at least one agentic system and 39\% experimenting, but the scaled deployments concentrate in digitally mature functions \citep{mckinsey2025stateofai}. BCG reported that while 72\% of workers use generative tools regularly, only 13\% of organizations have deployed agents integrated into broader workflows \citep{bcg2025aiatwork}. Gartner projects that over 40\% of agentic projects will be canceled by 2027, attributing the cancellations to unclear value and weak integration rather than to inadequate models \citep{gartner2025agentic}.

We propose that these numbers point to a missing abstraction rather than a missing capability. In operationally complex industries, value does not come from a single model invocation. It comes from coordinating fragmented tasks across tools, operators, documents, and approvals, while respecting constraints that cannot be violated and producing a record that operators can audit. This coordinating layer, which we call \emph{orchestration}, is largely absent from both the research conversation and the available tooling, and we believe its absence is the most coherent explanation for why capable models so often fail to become deployable systems.

This paper develops orchestration from a slogan into a frame one can think and design with. We make precise what it means for a workflow to be \emph{orchestration-bound} (Section~\ref{sec:gap}), decompose the gap into structural entanglement, coordination overhead, and recoverable value (Section~\ref{sec:framework}), and read three industries, logistics, healthcare operations, and construction, through that frame (Section~\ref{sec:cases}). We then state the requirements an orchestration runtime must meet, and where current frameworks fall short (Section~\ref{sec:arch}), and advance our central idea (Section~\ref{sec:staged}): automation should proceed in stages, and a sector's dominant friction determines which architectural guarantee is load-bearing. Section~\ref{sec:conclusion} lays out the research program this opens.

\section{The Orchestration Gap}
\label{sec:gap}

Model benchmarks measure something orthogonal to deployment success. An organization may have a model that drafts a contract or diagnoses a fault accurately and still be unable to deploy it, because the workflow infrastructure around the model does not exist. Orchestration is that infrastructure: the coordination of multi-step processes across many tools, actors, and decision points. The intuition is not new to those who build these systems. Fan et al. showed that workflow composition, parameter binding, and error handling are engineering problems distinct from language understanding \citep{fan2024workflowllm}, and surveys of multi-agent systems consistently report that the binding deployment obstacles are architectural rather than cognitive \citep{wu2023autogen}. What has been missing is a way to talk about this layer as a first-class object with its own structure and its own requirements.

\paragraph{Which workflows are orchestration-bound.} Not every workflow needs an orchestration layer; a single well-scoped model call suffices for many. The workflows that stall share a recognizable signature, which we capture in five properties, each stated as a checkable test:

\begin{enumerate}
\item \textbf{System fragmentation:} the workflow spans $\geq 3$ systems of record with no unifying platform.
\item \textbf{Stakeholder fragmentation:} execution requires handoffs among $\geq 3$ roles or organizations.
\item \textbf{Tacit decision points:} at least one consequential decision relies on uncodified judgment no written procedure fully specifies.
\item \textbf{Constraint criticality:} at least one hard constraint (regulatory, contractual, safety) must hold, and an output that violates it is unacceptable, not merely suboptimal.
\item \textbf{Brownfield pressure:} the workflow runs on entrenched legacy tools that operators would reject replacing wholesale.
\end{enumerate}

A workflow that satisfies at least four is orchestration-bound. The test makes the category checkable rather than rhetorical: one can disagree with a specific property for a specific workflow. All three focal industries satisfy all five, and the four further industries in Appendix~\ref{app:extended} also qualify.

\section{Decomposing Orchestration Value}
\label{sec:framework}

To reason clearly about where orchestration matters, it helps to separate three questions the literature tends to run together: how structurally tangled a workflow is, how much of its effort is coordination rather than substance, and how much that coordination is worth. We name these so they can be discussed independently.

\paragraph{Fragmentation ($F$).} Structural entanglement is driven by the \emph{interaction} of three counts, not their sum: the number of distinct systems $s$, of stakeholder roles $r$, and of cross-boundary handoffs $h$. A workflow with many systems but a single actor is far simpler than one with many of each, so we read entanglement multiplicatively, $F = \log_{10}(s \cdot r \cdot h)$, taking the logarithm so the quantity is additive in orders of magnitude and comparable across sectors of very different scale. $F$ is a property of the workflow, readable from a process map, and independent of any system one might deploy on it.

\paragraph{Coordination overhead ($C$).} The share of total effort spent on coordination, $C = T_{\text{coord}} / T_{\text{total}}$, is the portion an orchestration layer could in principle absorb. It is the variable most often directly visible in administrative-cost and time-and-motion studies.

\paragraph{Recoverable value ($V$).} For a workflow run $n$ times per year at coordination cost $c$ per run, with achievable reduction $\rho$, the recoverable value is $V = n \cdot c \cdot \rho$. The reduction $\rho$ is not a constant; it depends on how much autonomy the orchestration layer is granted, which is the subject of Section~\ref{sec:staged}.

The decomposition is conceptual, not arithmetic. High entanglement does not by itself imply high value: a tangled but low-volume workflow may not be worth automating, while a moderately tangled, high-volume one can carry enormous value. It also locates where evidence is firm ($C$, often measured) and where it is thin ($\rho$, rarely measured per stage), and separates what is intrinsic to a workflow ($F$, $C$) from what depends on automation ($\rho$), the boundary where the staged argument of Section~\ref{sec:staged} does its work.

\paragraph{Friction: the fourth consideration.} Entanglement and value describe the opportunity. They are silent on how hard it is to capture. We therefore track a fourth, deliberately qualitative axis: \emph{friction}, the regulatory, liability, and trust barriers that lengthen the path from opportunity to deployment. Friction does not change $V$; it governs how quickly $V$ can be realized and, as we argue later, which architectural guarantee a sector most needs. We keep friction qualitative because, unlike $F$ and $C$, it does not reduce to a count, and a manufactured score would imply a precision we do not have.

\section{Three Industries Through the Frame}
\label{sec:cases}

We read three industries through the decomposition, chosen to span the friction spectrum: low regulatory friction (logistics), high regulatory friction (healthcare operations), and high liability friction (construction). Table~\ref{tab:evidence} gathers the public evidence.

\begin{table*}[t]
\centering\small
\caption{Public evidence for the orchestration gap in three industries. $F$ inputs are counts for a representative end-to-end workflow; $C$ and $V$ inputs are drawn from the cited administrative-cost and productivity studies. The validated-effect row reports independently published automation results, which we use to anchor the reduction $\rho$.}
\label{tab:evidence}
\begin{tabularx}{\textwidth}{@{}lXXX@{}}
\toprule
& \textbf{Logistics (last-mile exceptions)} & \textbf{Healthcare ops (prior auth)} & \textbf{Construction (change orders)} \\
\midrule
Systems $s$ & 5--9 \citep{pournader2021aiscm} & 3--4: EHR, payer portal, scheduler, billing \citep{ehr2025scopingreview} & 4--6: model, schedule, cost, RFI \citep{mckinsey2020nextnormal} \\
Roles $r$ & 4--6 & 3--5: provider, staff, payer, patient & 5--8: GC, subs, architect, owner \\
Handoffs $h$ & 6--10 & 8--15 \citep{caqh2024index} & 10--15 per order \citep{mckinsey2020nextnormal} \\
\textbf{$F$} & \textbf{$\approx 5.6$} & \textbf{$\approx 5.1$} & \textbf{$\approx 5.9$} \\
\midrule
Coordination share & 53\% of shipping cost \citep{statista2023lastmile} & 34.2\% of spend is admin \citep{himmelstein2020admincosts} & rework 20\% of cost, 30\% of time \citep{mckinsey2017construction} \\
Per-unit cost & up to 26\% margin erosion \citep{capgemini2019lastmile} & \$3.41 / manual PA \citep{caqh2024index} & part of \$273B/yr error cost \citep{mckinsey2020nextnormal} \\
Validated effect (anchors $\rho$) & n/a & \$0.05 / electronic PA: 98\% cut \citep{caqh2024index} & n/a \\
Dominant friction & regulatory: low & regulatory: high & liability: high \\
\bottomrule
\end{tabularx}
\end{table*}

\paragraph{Three readings.} The three sit in a narrow fragmentation band ($F\approx5.1$ to $5.9$; Table~\ref{tab:evidence}) yet differ in where coordination cost concentrates and what gates its recovery. In logistics it concentrates in exception handling, where a failed delivery triggers a phone-coordinated cascade across routing, warehouse, customer, and carrier systems, and low regulatory friction lets routine exceptions reach bounded automation soonest. Healthcare administration gives the clearest evidence that the gap is orchestration rather than capability: a manual prior authorization costs \$3.41 against \$0.05 electronic, a 98\% reduction, yet only one in five is fully electronic \citep{caqh2024index}, so the records were digitized while the workflow joining them was not. Construction shows the largest coordination burden: rework runs about 20\% of cost and 30\% of time \citep{mckinsey2017construction}, the kind of waste a BIM-based coordination layer with clash detection is built to remove. The per-sector analyses, with workflow counts and friction readings, are in Appendix~\ref{app:cases}.

\paragraph{What the three together reveal.} The cases share a similar fragmentation band yet diverge in how automation should proceed, and the variable that explains the divergence is friction, not entanglement. This motivates our central claim. A theory of orchestration must do more than locate where coordination value is concentrated; it must say which guarantee a sector most needs, and that question is answered by friction. Figure~\ref{fig:groups} groups all eight industries (the focal three plus the four of Appendix~\ref{app:extended}) by friction along a shared fragmentation axis and makes the narrow band visible: structurally these workflows are alike, and friction is what separates them. That coordination is itself a large and addressable share of total effort is documented directly in Table~\ref{tab:evidence}: last-mile is 53\% of shipping cost, administration is 34\% of US health spending, and rework is 20\% of construction cost.

\begin{figure}[t]
\centering
\includegraphics[width=\columnwidth]{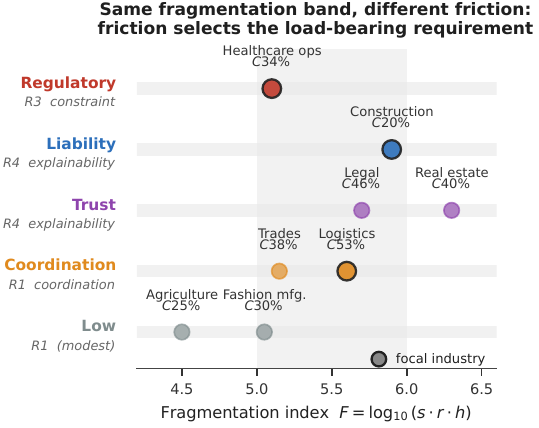}
\caption{The eight industries grouped by dominant friction and placed on a shared fragmentation axis $F$, with each industry's coordination overhead $C$ annotated and the focal three ringed. The shaded region marks the narrow band ($5.0\le F\le 6.0$) into which almost every industry falls, which is the point: structurally these workflows are alike, so $F$ flags orchestration-bound work but ranks it coarsely. Friction, not fragmentation, is what separates them, and it selects the load-bearing requirement named beneath each lane label. The $C$ values for the focal three are from cited sources (Table~\ref{tab:evidence}) and $F$ from cited workflow counts; the friction groupings are the authors' qualitative assessments.}
\label{fig:groups}
\end{figure}

\section{Requirements and Where Current Tools Stop}
\label{sec:arch}

\begin{figure*}[t]
\centering
\resizebox{0.60\textwidth}{!}{%
\begin{tikzpicture}[
  font=\sffamily\small,
  >={Stealth[length=2mm]},
  box/.style={rounded corners=2pt,draw,thick,minimum height=8mm,align=center,inner sep=3pt},
  lbl/.style={font=\sffamily\footnotesize\bfseries},
  flow/.style={->,thick,gray!70}
]
\node[box,fill=inputc!15,draw=inputc,minimum width=22mm] (ch1) {WhatsApp / Slack\\ Email / API};
\node[box,fill=inputc!15,draw=inputc,minimum width=22mm,below=2.5mm of ch1] (ch2) {Legacy systems\\ ERP / CRM / WMS};
\node[box,fill=inputc!15,draw=inputc,minimum width=22mm,below=2.5mm of ch2] (ch3) {Sensors / feeds\\ docs / forms};
\node[lbl,inputc,above=1mm of ch1] {Heterogeneous Inputs};
\node[box,fill=orchc!18,draw=orchc,minimum width=34mm,right=16mm of ch2] (router) {\textbf{Workflow Router}\\ \footnotesize multi-step coordination};
\node[box,fill=orchc!12,draw=orchc,minimum width=34mm,above=4mm of router] (state) {\textbf{State Manager}\\ \footnotesize long-running workflows};
\node[box,fill=orchc!12,draw=orchc,minimum width=34mm,below=4mm of router] (exec) {\textbf{Bounded Executor}\\ \footnotesize tool calls, error recovery};
\node[box,fill=knowc!15,draw=knowc,minimum width=30mm,above=14mm of state] (know) {\textbf{Domain Knowledge}\\ \footnotesize causal + constraint rules};
\node[box,fill=humanc!15,draw=humanc,minimum width=26mm,right=14mm of state] (human) {\textbf{Human Gates}\\ \footnotesize approve / modify / reject};
\node[box,fill=knowc!18,draw=knowc,minimum width=26mm,right=14mm of router] (constr) {\textbf{Constraint Gate}\\ \footnotesize block infeasible outputs};
\node[box,fill=outc!18,draw=outc,minimum width=26mm,right=14mm of exec] (out) {\textbf{Validated Output}\\ \footnotesize + audit trail};
\draw[flow] (ch1.east) -- (router.west);
\draw[flow] (ch2.east) -- (router.west);
\draw[flow] (ch3.east) -- (router.west);
\draw[flow] (know.south) -- (state.north);
\draw[flow,knowc] (know.east) to[out=0,in=180] (constr.north west);
\draw[flow] (state.south) -- (router.north);
\draw[flow] (router.south) -- (exec.north);
\draw[flow,humanc] (router.east) -- node[above,font=\tiny,black]{gated step} (human.west);
\draw[flow,humanc] (human.south) -- (constr.north);
\draw[flow,knowc] (exec.east) -- (constr.south west);
\draw[flow] (constr.south) -- (out.north);
\begin{scope}[on background layer]
\node[draw=orchc,dashed,thick,rounded corners=4pt,fill=orchc!4,
  fit=(state)(router)(exec),inner sep=6mm,label={[orchc,font=\sffamily\bfseries]above:Orchestration Layer (reference design)}] (orchbox) {};
\end{scope}
\node[box,fill=bggray,draw=gray,minimum width=20mm,below=20mm of ch3.west,anchor=west] (s1) {\footnotesize \textbf{Stage 1}\\ Observability};
\node[box,fill=bggray,draw=gray,minimum width=20mm,right=3mm of s1] (s2) {\footnotesize \textbf{Stage 2}\\ Recommend};
\node[box,fill=bggray,draw=gray,minimum width=20mm,right=3mm of s2] (s3) {\footnotesize \textbf{Stage 3}\\ Bounded auto};
\node[box,fill=bggray,draw=gray,minimum width=20mm,right=3mm of s3] (s4) {\footnotesize \textbf{Stage 4}\\ Supervised auto};
\draw[->,very thick,gray!80] (s1.east) -- (s2.west);
\draw[->,very thick,gray!80] (s2.east) -- (s3.west);
\draw[->,very thick,gray!80] (s3.east) -- (s4.west);
\node[font=\sffamily\footnotesize\itshape,below=1mm of s2.south east,anchor=north] {increasing autonomy, increasing trust requirement $\longrightarrow$};
\end{tikzpicture}%
}
\caption{A reference design for an orchestration layer, organized around the requirements of Section~\ref{sec:arch}. Heterogeneous inputs reach a workflow router backed by a state manager and a bounded executor. Domain knowledge grounds routing and feeds a constraint gate that blocks infeasible outputs. Human gates intercept decisions needing authorization, and every output carries an audit trail. The staged path (bottom) governs how much autonomy is granted.}
\label{fig:arch}
\end{figure*}
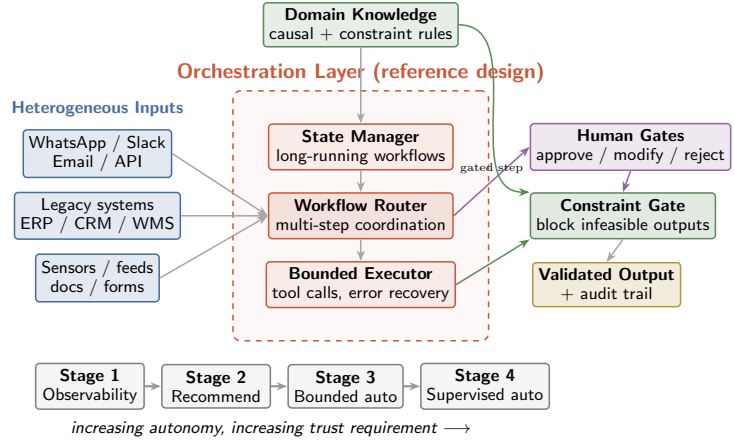

The cases imply a set of requirements any orchestration runtime must meet. Figure~\ref{fig:arch} sketches a reference design organized around them, and Table~\ref{tab:compare} reads current frameworks against the requirements as documented in their own materials.

\textbf{R1. Multi-step coordination with human gates:} run automated execution and human-authorized decisions in one workflow, escalating when approval stalls. \textbf{R2. Domain-knowledge grounding:} consult structured, queryable domain rules beyond what a general model carries. \textbf{R3. Constraint enforcement before delivery:} block constraint-violating outputs structurally, since an infeasible recommendation can be worse than none. \textbf{R4. Explainability:} emit a structured audit trace as a byproduct of execution. \textbf{R5. Brownfield integration:} interoperate with existing systems rather than replacing them.

\begin{table}[t]
\centering\small
\caption{Documented support for the five requirements across representative frameworks, per their public materials. \yes{} native, \parmark{} possible with engineering, \no{} not addressed.}
\label{tab:compare}
\setlength{\tabcolsep}{4.5pt}
\scalebox{0.95}{%
\begin{tabular}{@{}lccccc@{}}
\toprule
Framework & R1 & R2 & R3 & R4 & R5 \\
\midrule
AutoGen \citep{wu2023autogen} & \yes & \no & \parmark & \parmark & \parmark \\
MetaGPT \citep{hong2023metagpt} & \parmark & \no & \no & \parmark & \no \\
LangGraph \citep{langgraph2024} & \yes & \no & \parmark & \parmark & \parmark \\
\midrule
Requirement set & \yes & \yes & \yes & \yes & \yes \\
\bottomrule
\end{tabular}%
}
\end{table}

Table~\ref{tab:compare} captures the architectural claim. General multi-agent frameworks handle coordination (R1) well and offer human-in-the-loop hooks, but none treats domain-knowledge grounding (R2) or hard constraint enforcement (R3) as first-class, and brownfield integration (R5) is left to the implementer. None of this is unachievable on existing substrates: a constraint check can be coded as a graph node, and an interrupt can host a human gate. What operationally complex industries require is that these properties be \emph{architectural guarantees} the runtime enforces by construction, not optional patterns an integrator may or may not assemble correctly. That distinction, between a guarantee an operator can rely on for a regulated workflow and a convention they cannot, is what the deployment statistics in Section~\ref{sec:gap} reflect. Appendix~\ref{app:trace} walks a single prior authorization through this design from end to end.

\section{Staged Automation and the Friction Principle}
\label{sec:staged}

We now state our central idea. Automation in these industries should proceed in stages, and which architectural guarantee matters most is determined by a sector's dominant friction. Both halves of this claim are needed: the staging says \emph{how far} to go, and the friction principle says \emph{what to get right} along the way.

\begin{figure*}[!tb]
\centering
\includegraphics[width=0.92\textwidth]{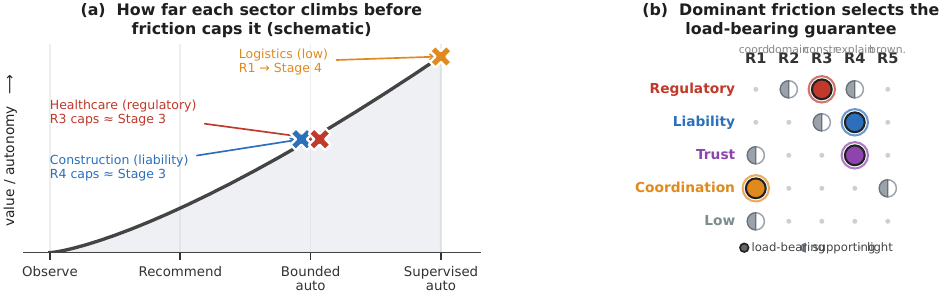}
\caption{The two halves of the central claim. \textbf{(a)} A schematic of the staged-automation path: the horizontal axis is the four stages of autonomy and the vertical axis the qualitative rise in recoverable value and granted autonomy, drawn without numeric ticks because the per-stage reduction $\rho$ is not measured; the marks ($\times$) show where each focal sector's dominant friction halts further autonomy. The endpoint is anchored by a published result, the CAQH electronic-PA reduction \citep{caqh2024index}; the curve shape and per-sector ceilings are the authors' qualitative reading. \textbf{(b)} The friction principle as a discrete map: each row is a friction profile and each column an architectural requirement (R1--R5); a filled ringed disk marks the load-bearing requirement, a half disk a supporting one, and a faint dot a light one. The levels are qualitative assessments grounded in the cited cases, not measured values.}
\label{fig:staged}
\end{figure*}

\paragraph{The stages.} The recoverable value $V = n\cdot c\cdot\rho$ rises with the granted reduction $\rho$, which in turn rises with autonomy. Four stages mark natural points on that curve. \textbf{Stage 1, observability} unifies workflow state across systems without deciding; its value is the elimination of status-chasing. \textbf{Stage 2, recommendation} proposes actions and flags exceptions while humans decide. \textbf{Stage 3, bounded automation} executes well-defined, low-risk subtasks autonomously. \textbf{Stage 4, supervised autonomy} executes multi-step workflows under human oversight rather than human initiation. The external evidence in Table~\ref{tab:evidence} anchors the endpoints of this curve: the CAQH manual-to-electronic transition shows a 98\% per-transaction reduction on qualifying prior-authorization subtasks \citep{caqh2024index}. As an illustration of scale, a clinic processing 6{,}000 prior authorizations per year at \$3.41 each spends roughly \$20{,}500 in direct coordination cost, of which a Stage-2 deployment might recover a fifth and a Stage-3 deployment half. Figure~\ref{fig:staged}(a) shows how far each focal sector climbs before friction caps it.

\paragraph{The friction principle.} The more important half of the claim is that the dominant friction in a sector selects which requirement from Section~\ref{sec:arch} is load-bearing. Under \emph{regulatory friction}, as in healthcare prior authorization, the binding requirement is constraint enforcement (R3): the value of the workflow is destroyed if an output violates payer or compliance rules, so the constraint gate, not raw throughput, is what makes the system deployable. Under \emph{liability friction}, as in construction change orders, the binding requirement is explainability (R4): because exposure is distributed across contractually independent parties, no party will accept an automated action unless the audit trail lets them verify it, so the trace is what makes the system deployable. Under \emph{low friction}, as in logistics exceptions, neither guarantee gates adoption, and a sector can move toward bounded automation primarily on the strength of coordination (R1) alone. The same orchestration value, in other words, calls for different architectural emphasis depending on friction, and a theory that ignored friction would prescribe the same system everywhere and be wrong in two cases out of three. Figure~\ref{fig:staged}(b) summarizes the principle: across friction profiles, a different architectural guarantee becomes load-bearing.

\paragraph{Why stage at all.} Trust and tuning accumulate: a runtime that has operated at Stage 2 carries a track record into Stage 3, while one that leaps to Stage 4 carries none. The thesis has boundaries as well, which we take up among the limitations.

\section{Conclusion and Research Program}
\label{sec:conclusion}

We have argued that process automation stalls in operationally complex industries because of a missing abstraction, orchestration, not a missing capability, and we developed that abstraction into a usable frame: a test for which workflows are orchestration-bound, a decomposition separating entanglement from coordination overhead from recoverable value, a feature-level account of where current frameworks stop, and a principle tying a sector's dominant friction to the guarantee it most needs.

This reorders priorities: the most useful near-term research is not another capability benchmark but the infrastructure to study orchestration directly. Three directions follow: workflow-level benchmarks that measure coordination cost, constraint-violation rate, and audit completeness rather than single-step accuracy; deployment studies that measure the reduction $\rho$ per stage on real workflows, turning the staged curve into a calibrated instrument; and head-to-head evaluation of runtimes on the five requirements, so the architectural-guarantee claim can be tested rather than argued. The orchestration gap is, we believe, both the reason for the present plateau and the most promising place to push next.

\clearpage
\section*{Limitations}

The frame we propose is built from public operational evidence rather than from a deployed system, and its constructs carry the corresponding caveats. The fragmentation quantity $F$ is intended to separate structural entanglement from value, not to rank workflows finely; in practice real industries cluster in a narrow band of $F$, and the quantity is sensitive to how a workflow's boundary is drawn, so it flags orchestration-bound workflows more reliably than it orders them. The reduction $\rho$ that drives recoverable value is anchored to independently published automation results rather than measured per stage on orchestration systems, which do not yet exist at the granularity our staging describes; the scale illustration in Section~\ref{sec:staged} should be read in that light. The framework comparison reflects documented capabilities rather than head-to-head benchmarks, and tooling evolves, so specific assessments will date. Our evidence is strongest for the three focal industries because public data is strongest there; the four further sectors in the appendix rest on industry reports of varying rigor. Finally, the staging thesis carries its own boundaries. We do not model deployment cost, which determines net rather than gross value: each stage carries integration and change-management cost, so for a low-volume workflow the early investment may never be recouped and the correct choice can be no orchestration at all. Orchestration is also necessary rather than sufficient, since a project that builds a coordination layer can still fail on data quality, incentives, or trust \citep{gartner2025agentic}; and human gates do not add value uniformly, helping on technical escalations more than on others \citep{wang2026hitl}, so they must sit where judgment is genuinely needed. Each of these is a direction for the research program above rather than a qualification we wish to leave unexamined.

\bibliography{refs}

\clearpage
\appendix

\section{Supplementary Material}
\label{app:main}

\subsection{Worked Trace}
\label{sec:trace}\label{app:trace}

To make the reference design concrete rather than schematic, we walk a single prior authorization through the components of the reference design (Section~\ref{sec:arch}). The example is illustrative, but every field and rule is the kind a real deployment would carry, and the trace shows how the five requirements act as guarantees rather than conventions. Table~\ref{tab:trace} summarizes the steps.

\begin{table}[t]
\centering\small
\caption{One prior authorization traced through the reference design. Each row names the component, the action, and the requirement (R1--R5) it realizes.}
\label{tab:trace}
\begin{tabularx}{\columnwidth}{@{}p{1.55cm}Xc@{}}
\toprule
Component & Action & Req. \\
\midrule
Router & Ingest order: \texttt{MRI lumbar, plan=BCBS-PPO, dx=M54.5} & R1 \\
Knowledge & Look up rule: this CPT under this plan \emph{requires} PA + 6wk conservative care & R2 \\
State & Open workflow; pull prior PT notes from EHR across 4 systems & R5 \\
Constraint & Check: documented PT $=$ 4wk $<$ 6wk $\Rightarrow$ \emph{block submission} & R3 \\
Human gate & Route to staff: ``insufficient conservative care; extend or override?'' & R1 \\
Constraint & After 6wk PT documented, re-check: rule satisfied & R3 \\
Executor & Submit electronically to payer portal; poll status & R5 \\
Trace & Log every step with timestamps and the rule version applied & R4 \\
\bottomrule
\end{tabularx}
\end{table}

The trace makes visible what prose about ``the system could help'' leaves abstract. The constraint gate (R3) is not advisory: it \emph{blocks} a submission that would be denied, before any staff time is spent, because the domain rule (R2) encodes the payer's six-week conservative-care requirement as a hard precondition rather than a hint. This is the behavior that makes the system deployable under regulatory friction, and it is exactly what a general framework leaves to the integrator. The human gate (R1) fires precisely where judgment is needed, at the decision to extend therapy or override, and nowhere else. And the audit trace (R4) records which rule version was applied, so a later denial or appeal can be reconstructed exactly. The same skeleton transfers across sectors: a construction change order would route the clash through the affected trades (R1), check the revised design against code constraints (R3), and log who approved what (R4), with explainability rather than constraint enforcement carrying the weight under liability friction.

\subsection{Extended Case Analyses}
\label{app:cases}

We give the full reading of the three focal industries summarized in Section~\ref{sec:cases}.

\paragraph{Logistics.} A shipment crosses order management, warehouse allocation, carrier selection, routing, dispatch, tracking, last-mile delivery, and exception handling, spanning five to nine disconnected platforms \citep{pournader2021aiscm}; with $s\!\approx\!7, r\!\approx\!5, h\!\approx\!8$, $F\approx5.6$. The coordination burden is unusually visible: last-mile delivery is 53\% of shipping cost \citep{statista2023lastmile}, and unoptimized operations erode margins by up to 26\% \citep{capgemini2019lastmile}. It concentrates in exception handling, where a failed delivery forces a phone-coordinated cascade across routing, warehouse, customer, and carrier systems. The orchestration opening is exactly this cascade: a runtime that ingests the failure, queries the affected systems, proposes a coordinated recovery, and executes it in one pass. Because regulatory friction is low, logistics can plausibly reach bounded automation of routine exceptions sooner than the other two sectors.

\paragraph{Healthcare operations.} We address administration only, not clinical care. US administrative cost is 34.2\% of health expenditure against 17.0\% in Canada \citep{himmelstein2020admincosts}. Prior authorization (PA) spans EHR, payer portal, scheduler, and billing ($s\!\approx\!4$) across provider, staff, payer, and patient ($r\!\approx\!4$) with 8--15 handoffs ($h\!\approx\!10$), giving $F\approx5.1$. Here the coordination cost is directly measured: a manual PA costs \$3.41 against \$0.05 electronic, a 98\% reduction, yet only one in five PAs are fully electronic \citep{caqh2024index}, while a single documentation task has been observed to require 346 clicks across 43 screens \citep{ehr2025scopingreview}. This is the clearest evidence anywhere that the gap is orchestration: the records were digitized, but the workflow connecting them was not, and the gap between the manual and connected path is an order of magnitude. High regulatory friction sets the ceiling: the constraint gate must check documentation against payer rules before submission, so autonomous submission is defensible only for clearly qualifying cases, with denials routed to people.

\paragraph{Construction.} A change order spans modeling tools, scheduler, cost system, and RFI tracker ($s\!\approx\!5$) across GC, subcontractors, architect, engineer, and owner ($r\!\approx\!6$), generating 10--15 actions per order ($h\!\approx\!12$), giving $F\approx5.9$. Construction labor productivity grew 1\% per year over two decades, with rework at 20\% of cost and 30\% of time, part of \$273B in annual error cost \citep{mckinsey2017construction,mckinsey2020nextnormal}. The rework a BIM-based coordination layer with clash detection targets is large: McKinsey puts construction rework at roughly 20\% of cost and 30\% of time \citep{mckinsey2017construction}. Liability friction shapes deployment differently from healthcare, because the barrier is distributed contractual exposure rather than a single regulator, so the explainable trace of who was notified and what was approved becomes the requirement that carries the most weight.

\subsection{Further Industries}
\label{app:extended}

We read four additional industries through the same frame. Public evidence is thinner here than for the focal three, so the readings are more approximate.

\paragraph{Fashion and Garment Manufacturing.}
The pre-production workflow spans design tools, PLM, factory systems, and messaging ($s\!\approx\!4$), across designer, pattern maker, merchandiser, and factory ($r\!\approx\!4$), with handoffs across three to five sample rounds plus material approvals ($h\!\approx\!10$): $F\approx5.1$. Each offshore sample round takes two to four weeks for shipping alone, and tacit pattern-making judgment resists codification. Friction is low to moderate and commercial rather than regulatory. The orchestration opening is the sample-revision loop: capturing fit notes, updating specifications, regenerating factory instructions, and triggering material reorders as one coordinated step.

\paragraph{Real Estate Operations.}
A residential transaction spans agent CRM, loan origination, title platform, and inspection reports ($s\!\approx\!4$), across six to eight parties ($r\!\approx\!7$), with 150--200 tasks and dozens of handoffs ($h\!\approx\!20{+}$): $F\approx6.3$, the highest in our analysis and a useful reminder that $F$ ranks coarsely. The lender-agent gap is costliest: underwriting conditions flow through email chains with no shared dashboard. NAR reports 79\% e-signature and 68\% generative-tool use, but fragmented across point solutions \citep{nar2025techsurvey}. Friction is moderate and jurisdiction-specific. The opening is transaction-lifecycle tracking that parses the contract for dates and contingencies and proactively coordinates the parties.

\paragraph{Trades Businesses.}
A service call spans intake, dispatch, mobile app, supply-house systems, municipal permit portals, and accounting ($s\!\approx\!5$), across customer, dispatcher, technician, and inspector ($r\!\approx\!4$), with handoffs through diagnosis, quote, parts, work, inspection, and invoice ($h\!\approx\!8$): $F\approx5.1$. Digital maturity is bimodal, and many small contractors run no app that integrates data. Friction is moderate, driven by municipal permitting. The opening is quote-to-completion, auto-populating permit requirements by jurisdiction and tracking inspections through to invoicing.

\paragraph{Legal Services.}
A corporate-transaction workflow spans document management, data room, e-signature, and billing ($s\!\approx\!4$), across partners, associates, paralegals, opposing counsel, and client ($r\!\approx\!5$), with a 50--200+ item closing checklist and dependent deadline chains ($h\!\approx\!15$): $F\approx5.7$. Deadlines form a dependency ecology with severe consequences for misses. Foundational work maps which legal tasks are amenable to automation \citep{surden2019ailawoverview,wolterskluwer2026}. Friction is high, driven by confidentiality and malpractice exposure, which makes explainability load-bearing much as in construction. The opening is matter-lifecycle management, beginning with deadline tracking before progressing to document generation.

\end{document}